\documentclass[9pt,twocolumn,twoside]{osajnl}
\journal{ol} % Choose journal (ao, aop, josaa, josab, ol, pr)

\setboolean{shortarticle}{true}
\usepackage{subfig}
\usepackage{amsmath}
\usepackage{balance}
\usepackage{ulem}
\usepackage{float}
\usepackage{placeins}
\titleformat{\paragraph}[runin]
  {\fontsize{10}{12}\sffamily\bfseries}
  {}
  {0em}
  {#1}

\title{Increased SNR in acousto-optic imaging via coded ultrasound transmission}

\author[1, *]{Ahiad Levi}
\author[1]{Sagi Monin}
\author[1]{Evgeny Hahamovich}
\author[2]{Aner Lev}
\author[2]{Bruno G. Sfez}
\author[1]{Amir Rosenthal} 

\affil[1]{Andrew and Erna Viterbi Faculty of Electrical Engineering, Technion – Israel Institute of Technology, Technion City 32000, Haifa, Israel}
\affil[2]{The Israel Center for Advanced Photonics (ICAP), Yavne 81800, Israel}

\affil[*]{Corresponding author: sahiadl@campus.technion.ac.il}

\begin{abstract}
Acousto-optic imaging (AOI) is a non-invasive method that uses acoustic modulation to map the light fluence inside biological tissue. In many AOI implementations, ultrasound pulses are used in a time-gated measurement to perform depth-resolved imaging without the need for mechanical scanning. However, to achieve high axial resolution, it is required that ultrasound pulses with few cycles are used, limiting the modulation strength. In this Letter, we developed a new approach to pulse-based AOI in which coded ultrasound transmission is used. In coded-transmission AOI (CT-AOI), one may achieve an axial resolution that corresponds to a single cycle, but with a signal-to-noise ratio (SNR) that scales as the square root of the number of cycles. Using CT-AOI with 79 cycles, we experimentally demonstrate over 4-fold increase in SNR in comparison to a single-cycle AOI scheme. 
\end{abstract}

\setboolean{displaycopyright}{false}

\begin{document}

\maketitle

One of the fundamental limitations of optical imaging of biological tissue is light scattering due to optical heterogeneity. 
At depths exceeding several transport lengths, scattering leads to the diffusion of light, which severely limits the imaging resolution that may be achieved \cite{durduran2010diffuse}. 
Additionally, optical imaging with diffused light often requires solving non-linear optimization problems in order to map tissue parameters.

Acousto-optic imaging (AOI) is a hybrid approach that overcomes the limitations of light diffusion by using acoustic modulation \cite{Marks1993}. 
Conventionally, AOI is performed by illuminating the tissue with a highly coherent continuous-wave (CW) laser and using ultrasound to locally modulate the  phase of the laser light inside the tissue. In AOI, the ultrasound-induced phase modulation is a result of two mechanisms \cite{Resink2012}: pressure-induce modulation of the refractive index and periodic movement of the optical scatterers. When the coherence length of the laser is sufficiently long, the local ultrasound-induced phase modulation inside the tissue is translated into an intensity modulation of the speckle pattern on the tissue boundary.
Thus, by measuring the modulation depth of the speckle on the tissue boundary it is possible to quantify the light fluence within the tissue at the positions in which the acoustic modulation was performed \cite{Leutz1995}.

AOI is capable of identifying both highly absorbing and highly scattering structures through their effect on the light fluence \cite{Wang2004}, facilitating applications such as early assessment of osteoporosis \cite{Lev2007}. Additionally, AOI can provide information on blood flow in the acoustically modulated regions through analysis of the spectral broadening of the speckle modulation \cite{Tsalach2015}.
While in most applications AOI is used as an independent technique for assessing tissue parameters, it may also be used as a complimentary technique to optoacoustic tomography (OAT). 
In previous works \cite{Daoudi2012, Hussain2018} it has been shown that the information provided by AOI can remove the bias in OAT images due to light attenuation, thus enabling OAT-image quantification

One of the major challenges of AOI is its low singal-to-noise ratio (SNR), which leads to long acquisition times \cite{Resink2012}. 
To increase the SNR and imaging speed, cameras are often used to process a large number of individual speckle grains in parallel \cite{Leveque:99, Li:02, Leveque-Fort2001}.
However, camera-based AOI is, in most implementations, incompatible with \textit{in vivo} imaging, in which speckle decorrelation occurs at rates that are faster than the acquisition rate of conventional cameras. 
To enable \textit{in vivo} imaging, sensitive wideband photodetectors are used to fully characterize the temporal behavior of a small number of speckles. An additional advantage of using fast detectors is that they enable performing the acoustic modulation with pulses \cite{Lev:00}. 
Using the time-of-flight principle, the modulated photons can be mapped at different depths without the need of mechanical scanning.
However, using acoustic pulses, rather than continuous acoustic modulation, effectively results in a weak modulation signal. 
While the modulation signal may be improved by increasing the pulse duration, this approach comes at the cost of reduced axial resolution.

\begin{figure}[b!]
    \vspace{-0.5cm}
    \centering
    \captionsetup[subfigure]{labelformat=empty}
    \subfloat{
        \includegraphics[height = 7.5cm]{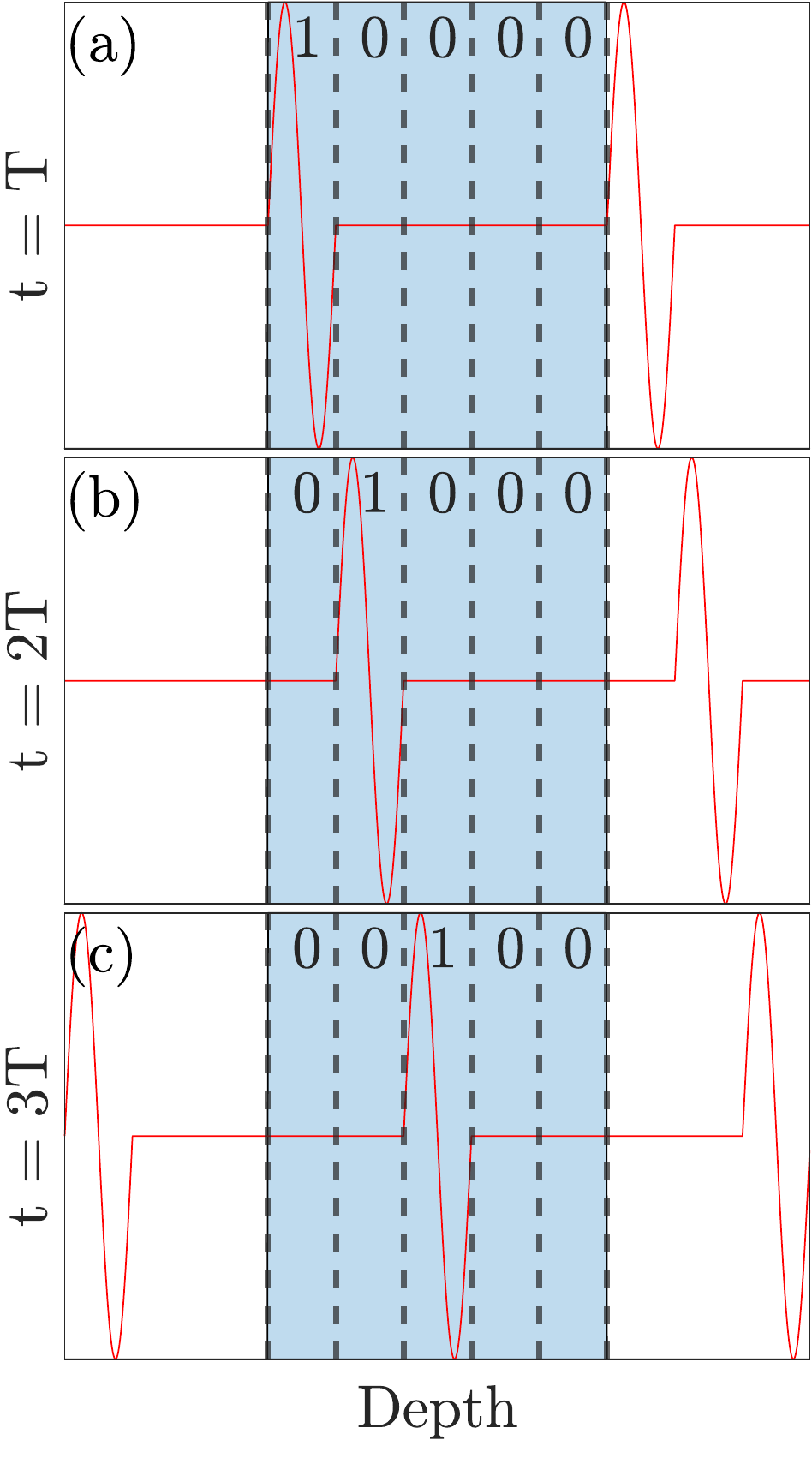}
        \label{fig:traditionalSim}
    } 
    %\hspace{0.01cm}%
    \subfloat{
        \includegraphics[height = 7.5cm]{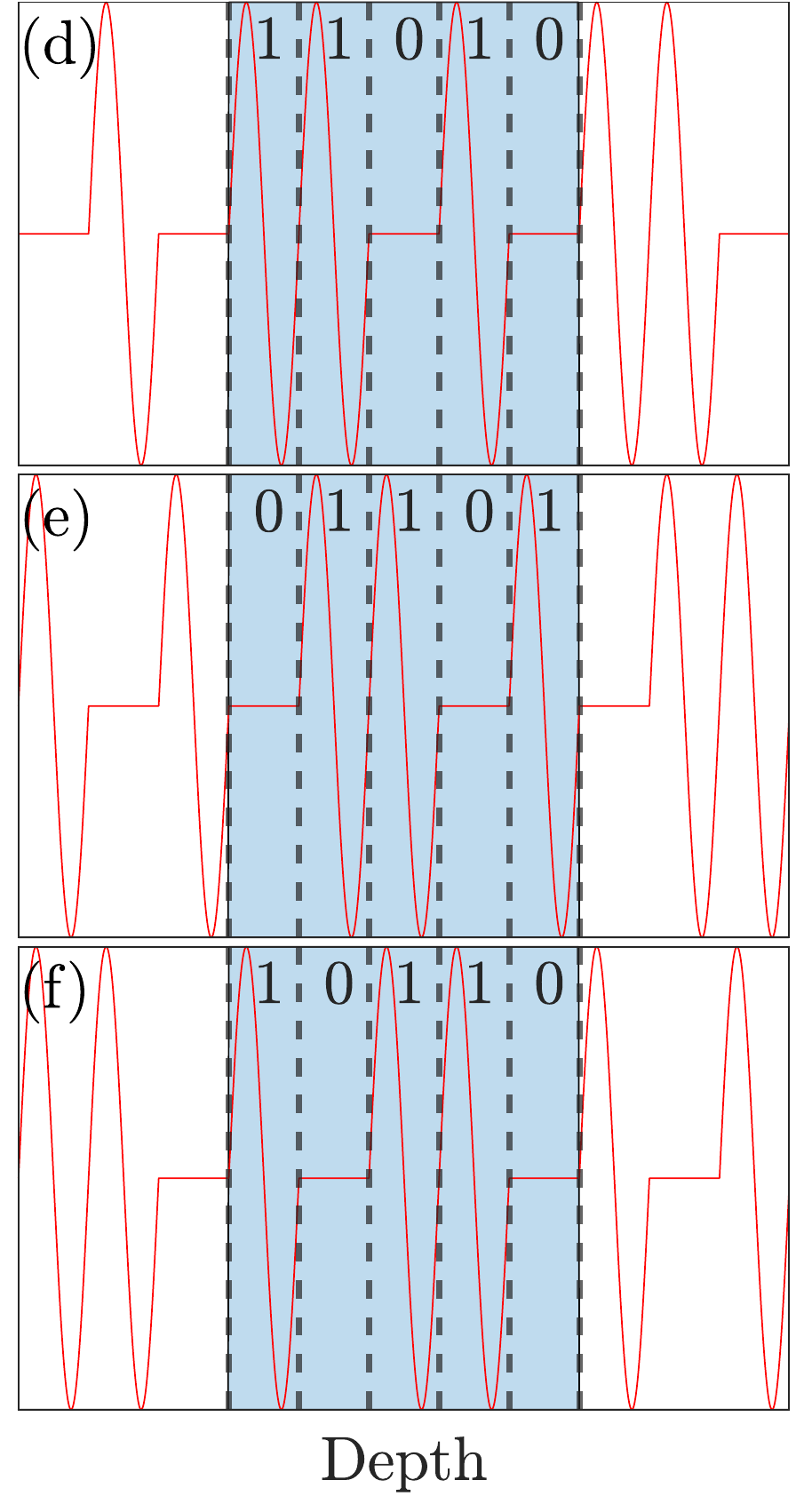}
        \label{fig:multiplexedSim}
    }
    % \vspace{-0.5cm}
    \caption{ (a-c) Traditional single-pulse AOI. Only one pulse is propagating through the phantom in a given time, modulating only the current region. (d-f) CT-AOI. A cyclic spatial combination of pulses modulating different regions inside the phantom along the propagation axis. Propagation of a single cyclic S-sequence of order $N$, is equivalent to multiplexing the tissue with $N$ different codes.
    }
    \label{fig:simulation}
    %\vspace{-0.5cm}
\end{figure}

In this work, we develop a new method to increase the SNR of pulse-based AOI without sacrificing the axial resolution. 
In our method, termed coded-transmission AOI (CT-AOI), a coded sequence of acoustic pulses is transmitted into the imaged object. 
Thus, at every time instance, the measured speckle modulation represents the sum of contributions from the acoustically modulated regions. 
Since the modulated regions change as the coded pulse train propagates, a set of multiplexed measurements is obtained, which may be used to recover the individual contribution from each region in the imaged medium. Our implementation was based on a cyclic sequence that achieves an SNR gain of $\sqrt{N}/2$, where $N$ denotes the number of elements in the code.

Fig. \ref{fig:simulation} shows an illustrative comparison between single-pulse AOI (Fig. \ref{fig:simulation}(a-c)) and CT-AOI (Figs. \ref{fig:simulation}(d-f)) in a 1D configuration. We divide the tissue into several sections, each with a width of $Tc$, where $T$ is the temporal width of a single-cycle pulse and $c$ is the speed of sound.
The acoustic pressure as a function of depth is represented by the red curve and by the corresponding binary values of the code shown above it. 
Accordingly, the 1s and 0s correspond to the modulated and unmodulated regions, respectively. 
The figure shows the propagation of the single pulse (Fig. \ref{fig:simulation}(a-c)) and pulse sequence  (Figs. \ref{fig:simulation}(d-f)) for three time instances. In the case of a single pulse, only one region in the medium is modulated at every time instance. In the case of a pulse sequence, the modulated regions are determined by the sequence, where at every time step $T$, the sequence is shifted cyclically. 

The code used in this work was an S-sequence, which is a known binary generalization of Hadamard codes in which the entries are either "0" or "1" \cite{harwit1979hadamard}. The code was generated as a Simplex code using the quadratic-residue method \cite{harwit1979hadamard} with a code order (length) of $N = 4m+3$, where $m$ is a natural number that leads to a value of $N$ that is a primary number. One of the properties of S-sequences is that they can be used to generate a basis in $\mathbf{R}^N$, where each basis vector is obtained by cyclic shifts of the S-sequence. Accordingly, we may construct an invertible matrix $\mathbf{S}$ whose first row is the S-sequence and every other row is circular shift of its preceding row. Representing the signal we wish to recover by the column vector $\mathbf{x}$, the multiplexed measurement $\mathbf{y}$ is obtained by $\mathbf{y}=\mathbf{Sx}$, and the recovery of $\mathbf{x}$ from the measurement can be performed by merely inverting the S-matrix:
\begin{equation}
    \label{eq:invert}
    \mathbf{X = S^{-1}Y}.
\end{equation}
where, the inverse of $\mathbf{S}$ was calculated numerically using the LU-decomposition method. We note that since $S$ is a well-conditioned square matrix, no generalized inversion procedure, e.g. using least-squares minimization, is required.

\begin{figure}[b!]
    \vspace{-0.5cm}
    \centering
    \includegraphics[width=\linewidth]{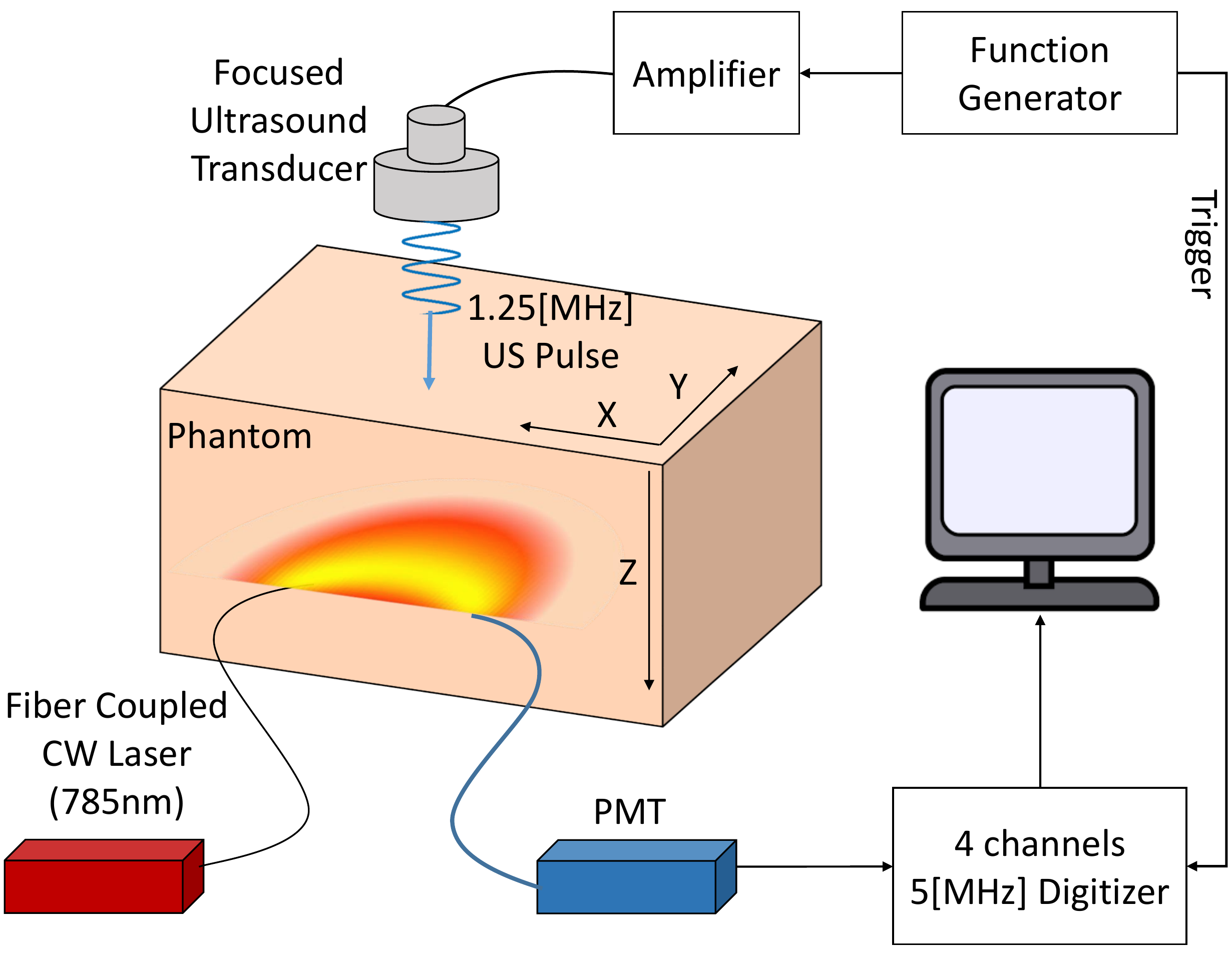}
    \vspace{-0.5cm}
    \caption{AOI system setup: An ultrasound pulse signal is generated using a function generator and electrically amplified before fed into the focused piezoelectric transducer. The pulse is propagting along Z axis into the illuminated phantom and modulating the tissue. Illumination is done by a 785nm fiber-coupled laser onto the phantom boundary. Sensing is done by a multimode fiber from the phantom boundary onto a photomultiplier tube. Both the illuminating fiber and the sensing fiber are placed on the same $XY$ plane. The signal is sampled at $f_s = 5MHz$ using a 4 channel digitizer, and then digitally processed to form an image.}
    \label{fig:systemSetup}
\end{figure}

\begin{figure*}[t!]
    \captionsetup[subfloat]{farskip=1pt,captionskip=2pt}
    \centering
    
    \subfloat{
        \includegraphics[height = 4cm]{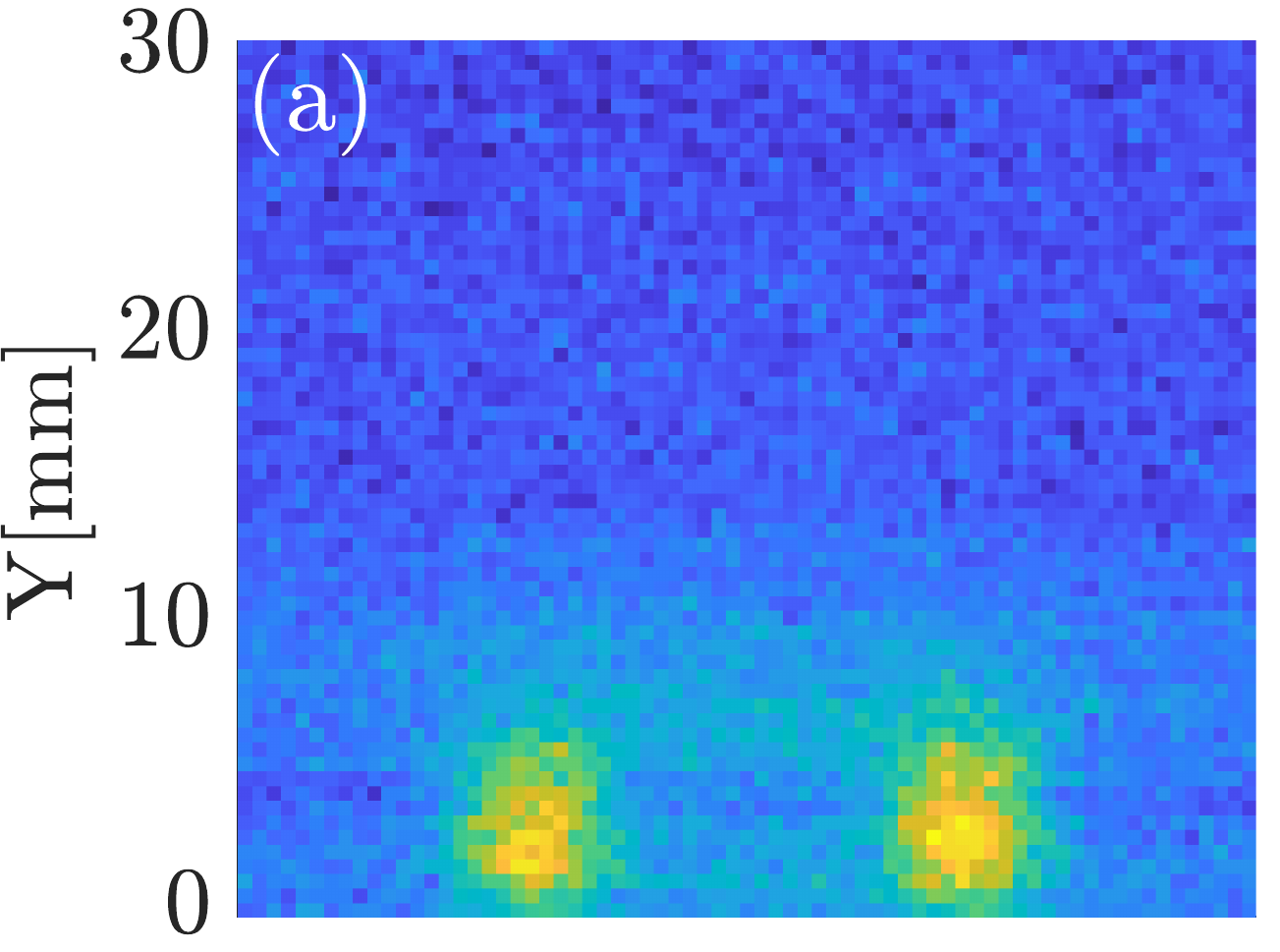}
        \label{fig:traditionalResLinear}
    }
    \subfloat{
        \includegraphics[height = 4cm]{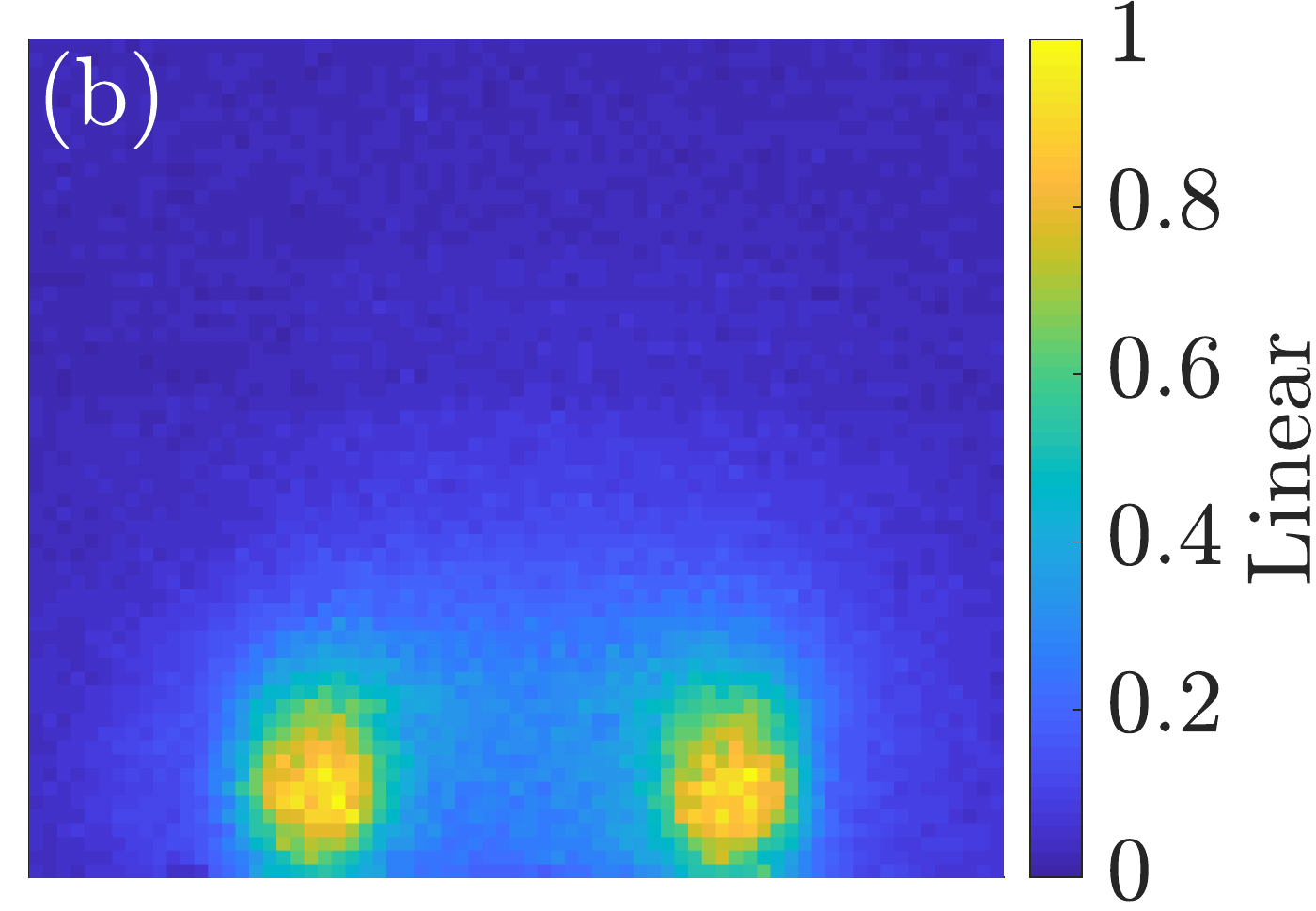}
        \label{fig:multiplexedResLinear}
    }
    
    \subfloat{
        \includegraphics[height = 4.94cm]{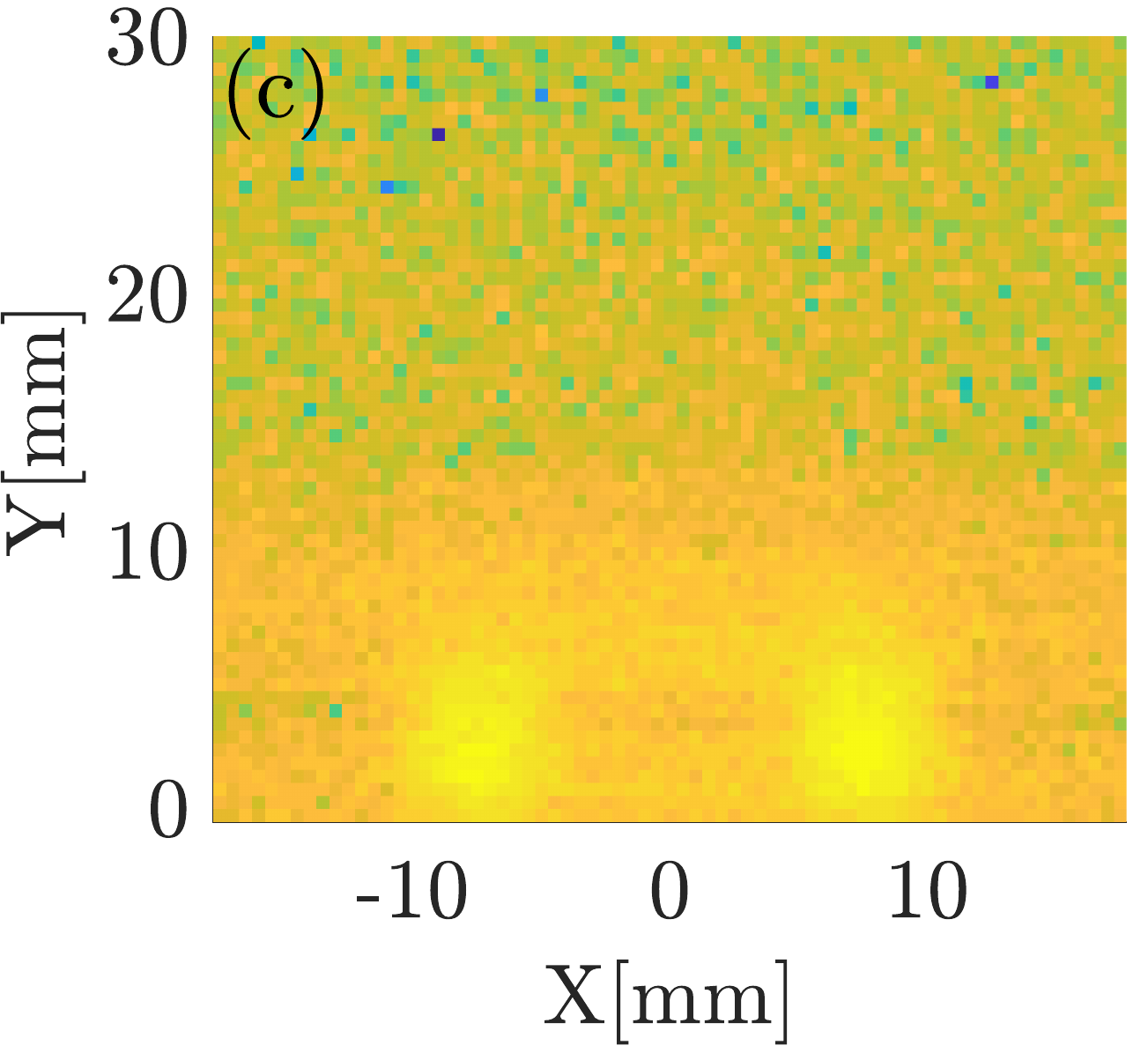}
        \label{fig:traditionalResDB}
    }
    \subfloat{
        \includegraphics[height = 4.95cm]{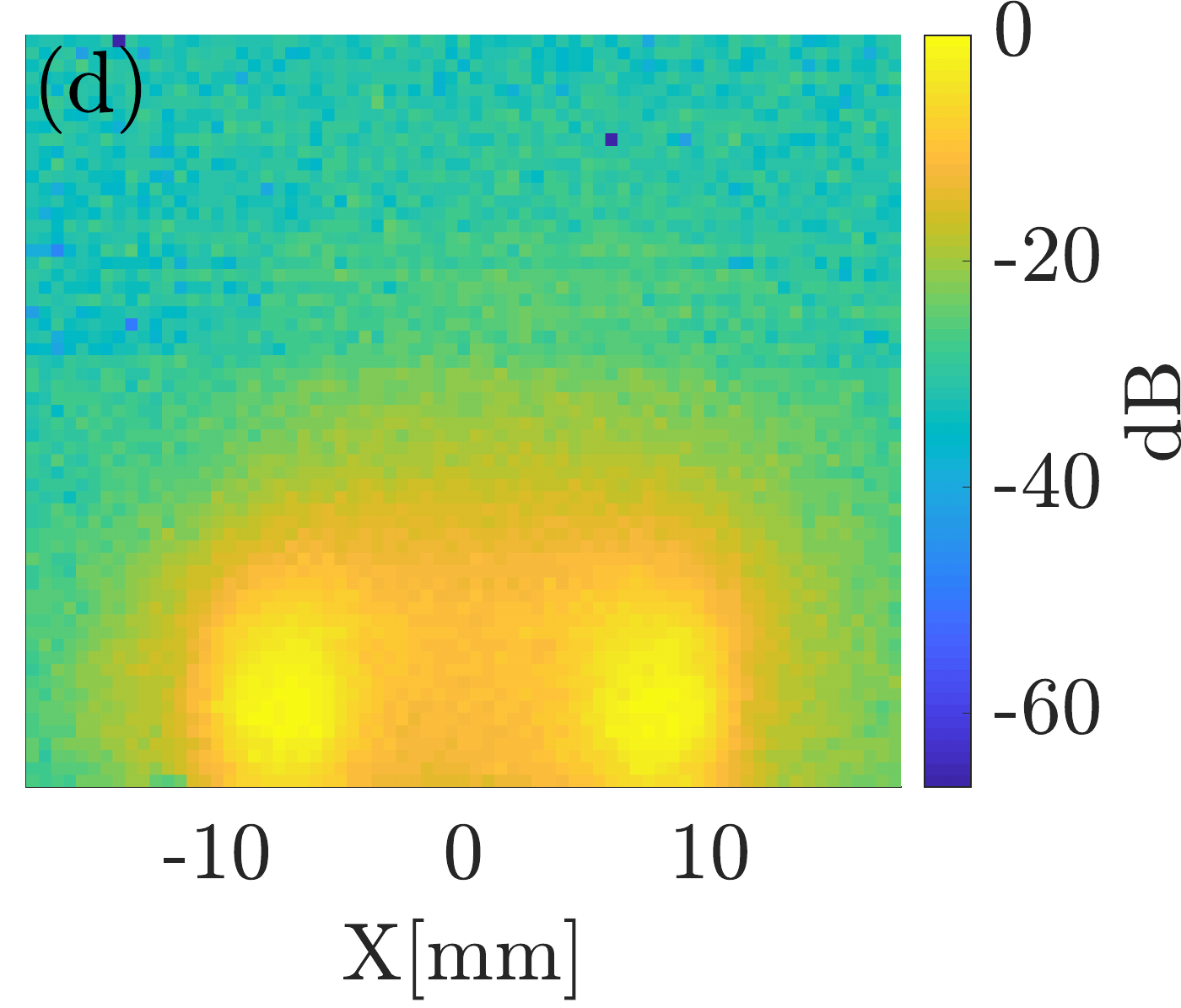}
        \label{fig:multiplexedResDB}
    }
    \caption{A comparison between fluence distribution inside the tissue mimicking phantom measured with single-pulse-AOI (a,c) and CT-AOI (b,d). The upper images presented in linear scale while the lower in logarithmic scale (dB). The CT-AOI approach introducing over a 4-fold in SNR, substantially eliminating the noise pattern from the image and yield a better SBR. The figure clearly shows that CT-AOI obtained a higher penetration than single-pulse-AOI. Quantitatively, the SNR in the single-pulse-AOI image dropped to 10 dB after only 7 mm, whereas for CT-AOI this SNR value was reached after 16.5 mm.}
    \label{fig:results}
\end{figure*}

In CT-AOI, since the shifts in the acoustic S-sequence delivered to the tissue are due to acoustic propagation, the result is continuous, rather than discrete, shifts. 
To enable a simple transition from the continuous to the discrete realms, the sampling frequency of the acoustically modulated light $f_s$, chosen as a multiple of the ultrasound frequency $f_{US}=T^{-1}$, i.e. the ratio $K=f_s/f_{US}$ is a natural number. 
Accordingly, the measured signal is divided into $K$ subsets, each with a sampling rate of $f_{US}$, where the relative shift between adjacent subsets is $T/K$. 
For each subset, the inversion is performed using Eq. \ref{eq:invert}, and the results are joined together into a single vector, representing the signal that would be obtained for a single-pulse US transmission with a sampling rate of $f_s$. 
To extract the modulated-portion of the demultiplexed signal, we use the algorithm developed in Ref. \cite{Lev:00}.

Figure \ref{fig:systemSetup} depicts the experimental setup of our CT-AOI system, which is based on design of Ref. \cite{Lev:03}. 
The illumination was provided by a CW laser (DL Pro 780, Toptica) with a wavelength of $785$ nm and linewidth of $50$ kHz. 
This wavelength is typical for deep-tissue-imaging applications since it lies in the near-infrared optical window, in which the optical absorption of blood and water are relatively low \cite{Yamada2014}. 
The laser was coupled to a multi-mode fiber with a diameter of 62.5 $\mu m$, where the power at the fiber output was 100 mW. 
The illumination was delivered  to an optically scattering phantom made  of clear silicone mixed with $193$ nm $\text{TiO}_2$ particles. 
The phantom had a reduced scattering coefficient of $\mu '_s = 15\;\text{cm}^{-1} $ and a speed of sound of approximately 990 m/s and was placed inside a water tank filled with distilled water. 
An additional multi-mode fiber with a diameter of 600 $\mu$m was used to collect the light scattered from the phantom and deliver it to a 4-channel photomultiplier tube (PMT, Hamamtsu). Both the illumination and collection fibers were on the same $X-Y$ plane, where the distance between the fibers was 15 mm. 
The electrical signal generated by the PMT was  sampled using a mutli-channel data-acquisition system (AlazarTech) at a sampling rate of $f_s = 5$ MS/s. 
The data acquisition was synchronized with the acoustic pulses by a trigger signal from the arbitrary function generator. 
In each acquisition sequence, the PMT signals were sampled over a duration of 2 s.

An arbitrary function generator (Tabor 8026) connected to a high voltage amplifier (.4-1.8-50EU26, SVPA) was used to generate high-voltage signals corresponding to single-pulse and coded transmission. 
As in Fig. \ref{fig:simulation}, each pulse contained exactly one cycle, where the central frequency was 1.25 MHz.
The voltage signals were fed to a focused transducer (Panametrics, A392S) with a diameter of 1.5 inch and focal length of 9.2 cm to generate an acoustic beam with a depth of field of 5.2 cm in the $z$ direction and a waist diameter of 1.8 mm.
The ultrasound transducer was positioned approximately 9 cm above the optical fibers used to illuminate the phantom and collect the scattered light. Two experiments were performed in this configuration.

\begin{figure}[b!]
    \vspace{-0.5 cm}
    \centering
    \includegraphics[width=\linewidth]{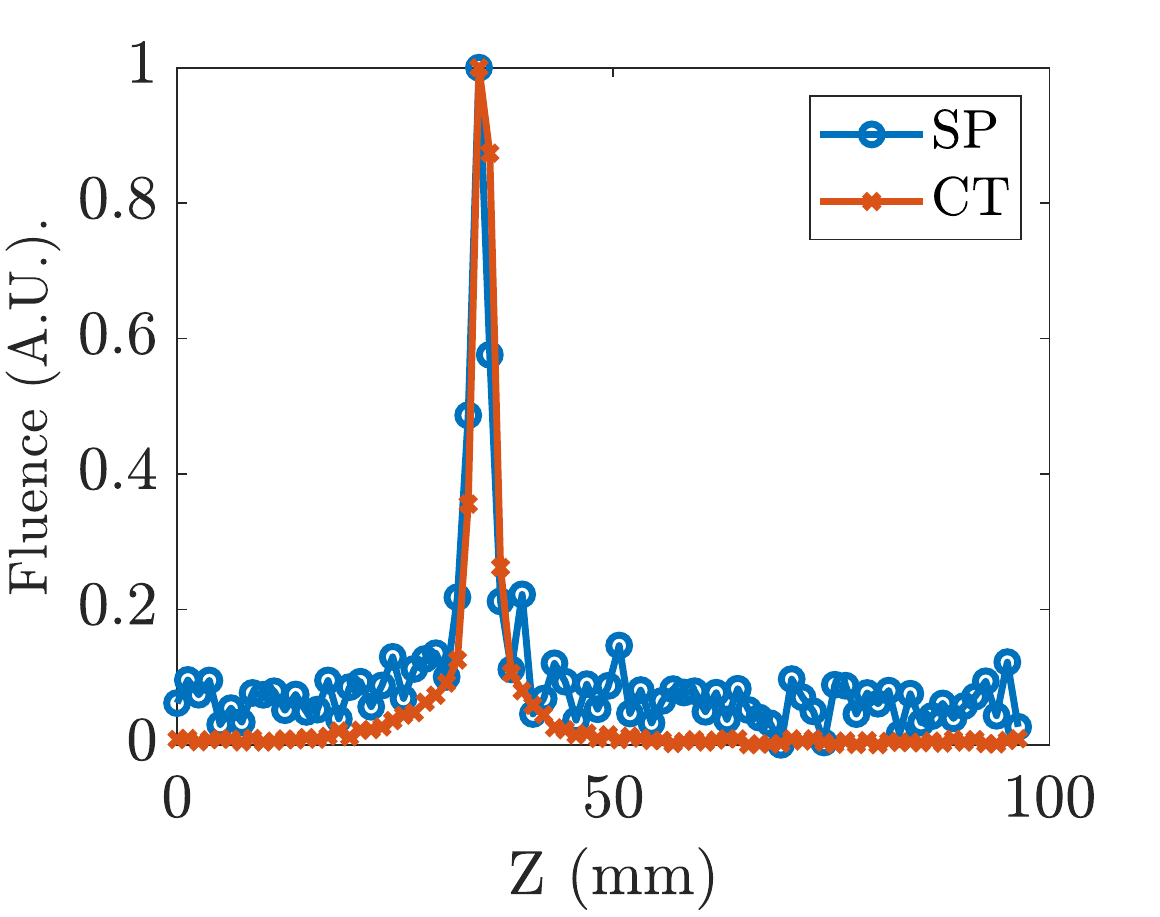}
    \vspace{-0.5cm}
    \caption{1D profile of the modulated light along the ultrasound propagation path for single-pulse AOI (blue) and CT-AOI (red) with corresponding FWMH values of 4.32 mm and 4.07 mm, respectively.} 
    \label{fig:reconstrcutionResolution}
    \vspace{-0.4 cm}
\end{figure}

In the first experiment, the light path from the illumination and collection fibers was measured. The ultrasound transducer was scanned in the $X$ and $Y$ directions with a step size of 0.5 mm. 
For a single-pulse transmission, a pulse-repetition rate of 15.82 kHz was set, corresponding to a distance of 9.4 cm between the pulses -- sufficiently long to assure no overlap between the pulses. 
When coded transmission was used, an S-sequence with $N=79$ was used, corresponding to the same propagation distance. Thus, at any given time, there was only a single S-sequence inside the phantom.

Figures \ref{fig:traditionalResLinear} and \ref{fig:multiplexedResLinear} show the normalized 2D maps of the modulated photons in the plane of the optical fibers obtained with single-pulse AOI and CT-AOI, respectively. 
In both cases, the well-known "banana-shape" light distribution from the source to the detector is obtained. However, the CT-AOI image achieved a 4.5-fold higher SNR, in agreement with the theoretical prediction. 
For comparison purposes, the images of modulated photons are also presented on a logarithmic scale in Figs. \ref{fig:traditionalResDB}  and \ref{fig:multiplexedResDB}.

The axial resolution in pulsed AOI is equal to the width of the ultrasound pulse used to modulate the light \cite{Lev:00}. 
Assuming an ideal transducer response for which each pulse contains a single cycle (Fig. \ref{fig:simulation}), a theoretical axial resolution of approximately 1.2 mm is obtained for our setup. 
Since the inversion procedure in Eq. \ref{eq:invert} is exact, this value is true for both single-pulse AOI and CT-AOI. 
To experimentally demonstrate that the use of coded pulses does not degrade the axial resolution, we compared the depth profiles (Z-dependence) of the AOI images at the X-Y position where the peak values were achieved in the 2D scan (Fig. \ref{fig:results}).
Figure \ref{fig:reconstrcutionResolution} compares the two depth profiles obtained for single-pulse AOI (blue) and CT-AOI (red); the corresponding FWHM values obtained from these data were 4.32 mm and 4.07 mm, respectively.

\begin{figure}[h!]
    % \vspace{-0.4 cm}
    \centering
    \includegraphics[width=\linewidth]{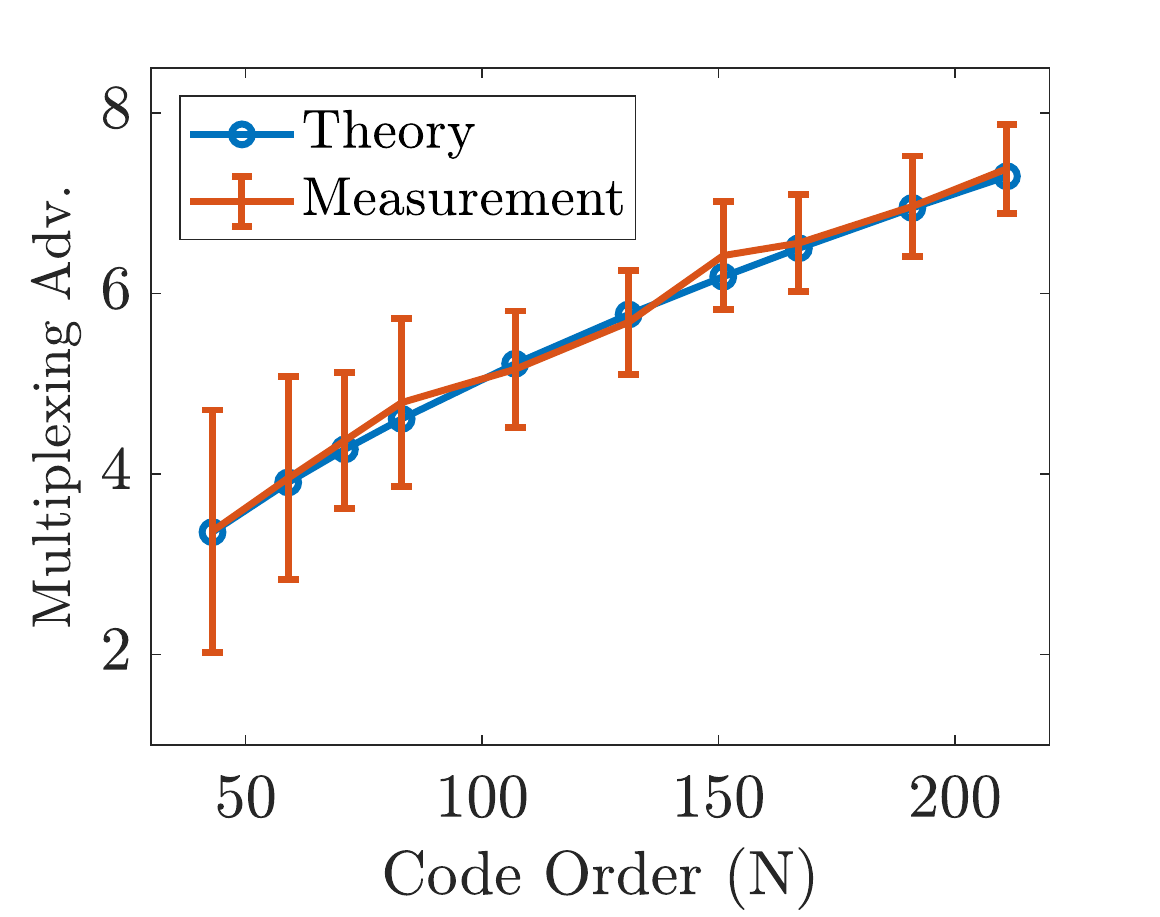}
    \vspace{-0.5cm}
    \caption{Multiplexing advantage: In blue, the theoretical value calculated by $\sqrt{N}/2$. In red experimental results measured in the discussed CT-AOI system. During the measurement, pulses were practically 1 cycles of sine function for both single-pulse AOI and CT-AOI.}
    \label{fig:fellgettAdvantage}
\end{figure}

In the second experiment, the multiplexing advantage was measured for different code orders with relation to its corresponding single pulse repetition rate. 
The transducer was positioned above the point in the $X-Y$ plane for which the highest signal value was obtained in the first experiment (Fig. \ref{fig:multiplexedResLinear}).
Coded ultrasound sequences with varying code lengths and single-pulse ultrasound were delivered to the phantom and the signal was measured over 2 s. 
To assess the SNR for each case, the measurement was repeated 30 times, and the average values of the signal and noise values were calculated. 
The multiplexing advantage was calculated by dividing the average SNR obtained for each code length by the SNR of the single-pulse measurement.

\balance

The measured values of the multiplexing advantage are presented in Fig. \ref{fig:fellgettAdvantage} (red curve) and are compared to the theoretical prediction of a multiplexing advantage of $\sqrt{N}/2$ (blue curve). 
The figure clearly shows a good agreement between the experimental results and the theoretical prediction.  
It is important to note that the SNR gain shown in Fig. \ref{fig:fellgettAdvantage} of CT-AOI represents a comparison between the SNR obtained for a single coded sequence and a single pulse. 
This type of comparison does not take into account that the repetition rate of the single pulse can be increased as long as there is only a single pulse inside the tissue, enabling better SNR via averaging. 
Thus, the SNR gain from very long sequences, e.g. $N>100$ in Fig. \ref{fig:fellgettAdvantage}, is obtained only when the repetition rate of single-pulse AOI is artificially reduced below its maximum value.

In conclusion, we introduced a new approach for increasing the SNR in pulse-based AOI, which relies on coded ultrasound transmission. 
Using an S-sequence code of the order of 79, a 4.5-fold SNR gain was achieved with CT-AOI over single-pulse AOI, in agreement with multiplexing theory. 
We note that this gain in sensitivity was achieved without sacrificing the longitudinal resolution or increasing the measurement duration. 
This property makes CT-AOI a promising approach for increasing the penetration depth of AOI in future \textit{in vivo} applications such as early detection of disease based on changes in the optical properties of the affected tissue \cite{Lev2007}, blood-flow estimation \cite{Tsalach2015}, and integration with clinical optoacoustic systems \cite{Daoudi2012, Hussain2018}.

\paragraph{Funding.}
This work has received funding from the Ministry of Science, Technology and Space (3-12970) and from the Ollendorff Minerva Center.

\paragraph{Disclosures.}
The authors declare no conflicts of interest.

% Bibliography
\bibliography{sample}
% Full bibliography added automatically for Optics Letters submissions; the following line will simply be ignored if submitting to other journals.
% Note that this extra page will not count against page length
\bibliographyfullrefs{sample}

\end{document}